\documentclass[twoside,twocolumn,english,pra,twocolumn,aps,superscriptaddress,showpacs]{revtex4-1}

\usepackage[T1]{fontenc}
\usepackage[latin9]{inputenc}
\usepackage{color}
\usepackage{babel}
\usepackage{bm}
\usepackage{amsmath}
\usepackage{amsthm}
\usepackage{amssymb}
\usepackage{graphicx}
\usepackage{esint}
\usepackage[unicode=true,pdfusetitle,
 bookmarks=true,bookmarksnumbered=false,bookmarksopen=false,
 breaklinks=false,pdfborder={0 0 0},backref=false,colorlinks=true]
 {hyperref}
\usepackage{breakurl}

\makeatletter
\theoremstyle{plain}

\usepackage{tabularx}

\usepackage{makecell}
\usepackage{fontenc}
\newcommand{\tabincell}[2]{\begin{tabular}{@{}#1@{}}#2\end{tabular}}
\hypersetup{urlcolor=blue}

\makeatother

\begin{document}

\preprint{This line only printed with preprint option}

\title{Optimal Conventional Measurements for Quantum-Enhanced Interferometry}

\author{Wei Zhong}

\email{zhongwei1118@gmail.com}

\selectlanguage{english}%

\affiliation{Guangdong Provincial Key Laboratory of Quantum Engineering and Quantum
Materials, SPTE, South China Normal University, Guangzhou 510006,
China}

\author{Yixiao Huang}

\affiliation{School of Science, Zhejiang University of Science and Technology,
Hangzhou 310036, China}

\author{Xiaoguang Wang}

\email{xgwang@zimp.zju.edu.cn}

\selectlanguage{english}%

\affiliation{Zhejiang Institute of Modern Physics, Department of Physics, Zhejiang
University, Hangzhou 310027, China}

\author{Shi-Liang Zhu}

\email{slzhu@nju.edu.cn}

\selectlanguage{english}%

\affiliation{National Laboratory of Solid State Microstructures and School of
Physics, Nanjing University, Nanjing 210093, China}

\affiliation{Guangdong Provincial Key Laboratory of Quantum Engineering and Quantum
Materials, SPTE, South China Normal University, Guangzhou 510006,
China}

\affiliation{Synergetic Innovation Center of Quantum Information and Quantum Physics,
University of Science and Technology of China, Hefei, Anhui 230026,
China}
\begin{abstract}
A major obstacle to attain the fundamental precision limit of the
phase estimation in an interferometry is the identification and implementation
of the optimal measurement. Here we demonstrate that this can be accomplished
by the use of three conventional measurements among interferometers
with Bayesian estimation techniques. Conditions that hold for the
precision limit to be attained with these measurements are obtained
by explicitly calculating the Fisher information. Remarkably, these
conditions are naturally satisfied in most interferometric experiments.
We apply our results to an experiment of atomic spectroscopy and examine
robustness of phase sensitivity for the two-axis counter-twisted state
suffering from detection noise.
\end{abstract}
\maketitle

\emph{Introduction.---}Quantum-enhanced interferometry has attracted
considerable attentions due to possible suppressing the uncertainty
of the phase measurement below the shot noise limit with quantum resources
like squeezing and entanglement \citep{Caves1981PRD,Yurke1986PRA,Wineland1992PRA,Holland1993PRL,Dowling1998PRA,Giovannetti2004Sci,Giovannetti2006PRL,Giovannetti2011nph}.
This quantum enhancement has potential application on significantly
improving weak signal detection and atomic frequency measurement.
Theoretically, the phase measurement sensitivity is limited by quantum
Cram\'er-Rao bound (QCRB) \citep{Helstrom1976Book,Holevo1982Book,Braunstein1994PRL},
which crucially depends on both the property of the probe state and
the way of phase accumulation. One practical difficulty in this field
is that the optimal measurements to access this theoretical precision
limit are often not physically realizable \citep{Braunstein1994PRL,Durkin2010NJP,Zhong2014JPA}.
More recently, some novel detection methods, such as single-particle
detection \citep{Zhang2012PRL,Hume2013PRL,Monras2006PRA}, interaction-based
detection \citep{Davis2016PRL,Froewis2016PRL,Macr`i2016PRA}, and
weak measurement \citep{Hofmann2011PRA,Hofmann2012PRA,Pang2015PRL,Zhang2015PRL}
were raised on some specific problems of phase estimation. Whereas,
more popular detection methods used in interferometric phase measurement
are those relative to the particle count or the population difference
on the output ports of the interferometer, since they are readily
implementable with current experimental techniques. A question of
primary concern, therefore, is whether the fundamental precision limit
can be attained by these conventional measurements. 

Recently, some progress has been made in this aspect. In Ref.~\citep{Pezze2008PRL},
Pezz\'e and Smerzi first simulated a Bayesian analysis in optical
interferometry that the QCRB can be saturated by the two-output-port
(TOP) photon count measurement for the state created by the interference
between squeezing laser and coherent laser. Hofmann subsequently proved
that this result can be generalized to all path-symmetric pure states
\citep{Hofmann2009PRA}. In another recent work ~\citep{Pezze2013PRL},
Pezz\'e and Smerzi analytically found that the single-output-port
(SOP) photon count measurement can attain the QCRB for single-mode
number squeezing states at a particular value of the phase shift $\theta\!=\!0$.
Furthermore, there were some experimental evidences indicating that
for certain typical quantum states even a population difference (PD)
measurement is sufficient with the Bayesian analysis \citep{Krischek2011PRL,Strobel2014Sci}.
However, it is still ambiguous for the roles of these conventional
measurements in precision measurement. For instance, in what circumstances
and under what conditions can these measurements implemented achieve
the quantum Cram\'er-Rao sensitivity? What differences are among
these measures?

In this manuscript, we address these issues by explicitly calculating
the Fisher information under these three measurements for a general
class of quantum states in the interferometry. We clarify that all
these measurements are conditionally optimal for achieving the ultimate
sensitivity. Specifically, when the probe state---the state before
the phase shift operation---be a real symmetric pure state, the QCRB
can be reached by counting the particle number of the two output ports
of the interferometer in the whole phase interval without overhead
of extra feedback operations \citep{Sanders1995PRL,Berry2000PRL,Berry2009PRA}.
This can be equivalently accomplished by measuring the population
difference between the two output states when probe states chosen
are absence of the fluctuating particle number. While the measurement
of counting particle number on a single interferometer output only
saturates for the particular values of the phase shift $\theta\!=\!0$
and $\pi$. As for the cases of complex symmetric pure states, all
three measures share the same performance such that the QCRB can be
attained at $\theta\!=\!0$ and $\pi$. More friendly, all these requirements
are readily met in current experiments on high precision phase estimation,
since most of quantum states employed in experiments belongs to the
family of symmetric pure states, for instance, states created by the
interference between an arbitrary state and an even or odd state \citep{Caves1981PRD,Holland1993PRL,Pezze2008PRL,Afek2010Sci,Joo2011PRL,Krischek2011PRL,Liu2013PRA,Pezze2013PRL,Lang2013PRL}
and two-mode squeezed vacuum state \citep{Anisimov2010PRL} used in
optical settings, as well as squeezed spin states \citep{Wineland1992PRA,Meyer2001PRL,Pezze2009PRL,Strobel2014Sci}
and atomic entangled states \citep{Bollinger1996PRA,Leibfried2004SCI,Lucke2011Sci}
used in atomic settings. In addition, these results can be readily
generalized into case of absence of phase reference frame accompanying
with probe states being phase averaged \citep{Bartlett2007RMP,Hyllus2010PRL,Jarzyna2012PRA}. 

\begin{figure}[t]
\includegraphics[scale=0.55]{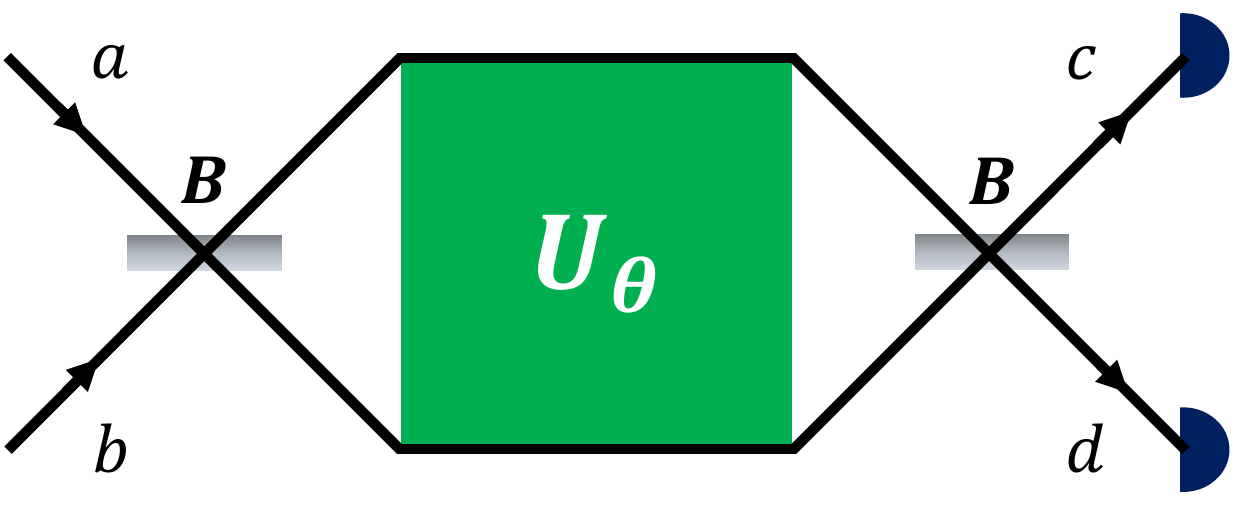}\caption{(Color online) Schematic of the standard MZI. \label{fig:MZI}}
\end{figure}

\emph{Mach-Zehnder interfermometer (MZI).---}We start by modeling
the setup of interferometric phase measurement onto a linear MZI (see
Fig.~\ref{fig:MZI}). The routine means for studying two-mode interferometer
is by using the Schwinger representation of the angular momentum operators
$\hat{J}_{x}\!=\!(\hat{a}^{\dagger}\hat{b}\!+\!\hat{b}^{\dagger}\hat{a})/2$,
$\hat{J}_{y}\!=\!(\hat{a}^{\dagger}\hat{b}\!-\!\hat{b}^{\dagger}\hat{a})/2i$,
and $\hat{J}_{z}\!=\!(\hat{a}^{\dagger}\hat{a}\!-\!\hat{b}^{\dagger}\hat{b})/2$,
where $\hat{a}\thinspace(\hat{a}^{\dagger})$ and $\hat{b}\thinspace(\hat{b}^{\dagger})$
are the annihilation (creation) operators of the upper and lower input
modes of the interferometer, respectively. These operators commute
with the total particle number operator $\hat{N}\!=\!\hat{a}^{\dagger}\hat{a}\!+\!\hat{b}^{\dagger}\hat{b}$
and satisfy the commutation relations for Lie algebra $\mathfrak{su}\left(2\right)$.
A standard MZI is made up of two $50\!:\!50$ beam splitters $B\!=\!\exp(-i\pi\hat{J}_{y}/2)$
and a phase shift $U_{\theta}\!=\!\exp(-i\theta\hat{J}_{z})$ in terms
of an unknown $\theta$ to be estimated. Let $\vert\psi_{{\rm in}}\rangle$
denote the state entering at the input ports of the interferometer.
Then the state at the ouput ports reads $\vert\psi_{{\rm out}}\rangle\!=\!B^{\dagger}\thinspace\!U_{\theta}B\vert\psi_{{\rm in}}\rangle$. 

To facilitate our analysis, we focus on the state after the first
beam splitter $B\vert\psi_{{\rm in}}\rangle$, which is called as
probe state and denoted by $\vert\psi\rangle$. We use the basis space
spanned by common eigenstates $\vert j,m\rangle$ of the operators
$\hat{\bm{J}}^{2}\!\equiv\!\hat{J}_{x}^{2}+\hat{J}_{y}^{2}+\hat{J}_{z}^{2}$
and $\hat{J}_{z}$ with eigenvalues $j\left(j+1\right)$ and $m$,
respectively. An arbitrary pure state $\vert\psi\rangle$ can be generally
expressed as $\vert\psi\rangle\!=\!\sum_{j}\sum_{m=-j}^{j}C_{j,m}\vert j,m\rangle$
where $C_{j,m}$ denote the expansion coefficients of $\vert\psi\rangle$
on $\vert j,m\rangle$. We here restrict our attention to a family
of symmetric pure states with the expansion coefficients satisfying
$C_{j,m}\!=\!C_{j,-m}$, suggesting $\langle\hat{J}_{z}\rangle_{\psi}\!=\!0$.
Notice that such a family covers a wide range of quantum states as
mentioned previously.

According to quantum estimation theory, the QCRB states that, for
a given parametric density matrix $\rho_{\theta}$, the phase sensitivity
of any unbiased estimator is bounded by $\Delta\theta_{{\rm est}}^{2}\!\leq\![\upsilon F_{Q}(\rho_{\theta})]^{-1}$
where $\upsilon$ is the number of independent measurements and $F_{Q}(\rho_{\theta})$
is the so-called quantum Fisher information (QFI) \citep{Helstrom1976Book,Holevo1982Book,Braunstein1994PRL}.
In the case as described above, the QFI in terms of the phase-imprinted
symmetric pure state $\vert\psi_{\theta}\rangle\!\equiv\!U_{\theta}\vert\psi\rangle$
is exactly given by

\begin{equation}
F_{Q}(\psi_{\theta})=4\langle\hat{J}_{z}^{2}\rangle_{\psi}=8\sum_{j}\!\sum_{m=j-\left\lfloor j\right\rfloor }^{j}\!\vert C_{j,m}\vert^{2}m^{2},\label{eq:QFI_Jz}
\end{equation}
with $\left\lfloor \thinspace\thinspace\right\rfloor $ denoting the
corresponding integer part. Remarkably, the expression in Eq.~\eqref{eq:QFI_Jz}
is irrelevant to the phase parameter $\theta$, meaning that the ultimate
sensitivities provided by symmetric pure states do not depend on the
true value of the phase shift.

In experiments to achieve the QCRB, one needs to successively perform
optimizations over measurements and estimators. A generic optimization
procedure is as follows. Consider a general measurement described
by a positive-operator-valued measure $\hat{\bm{M}}\!\equiv\!\{\hat{M}_{\chi}\}$
with $\chi$ the results of measurement. Based on $\hat{\bm{M}}$,
the accessible phase sensitivity is limited by the inequality $\Delta\theta_{{\rm est}}^{2}\!\leq\![\upsilon F(\rho_{\theta}\vert\hat{\bm{M}})]^{-1}$,
where $F(\rho_{\theta}\vert\hat{\bm{M}})$ is the classical Fisher
information (CFI) defined from below as 
\begin{equation}
F(\rho_{\theta}\vert\hat{\bm{M}})=\sum_{\chi}\frac{1}{p(\chi\vert\theta)}\bigg(\frac{d\,p(\chi\vert\theta)}{d\theta}\bigg)^{2},\label{eq:CFI}
\end{equation}
with $p(\chi\vert\theta)\!\equiv\!{\rm Tr}(\hat{M}_{\chi}\rho_{\theta})$
the probability of the outcome $\chi$ conditioned on the specific
value of $\theta$. It is well known that such accessible sensitivity
bound is achieved by the maximum likelihood estimator for sufficiently
large $\upsilon$ with Bayesian estimation methods \citep{Uys2007PRA,Krischek2011PRL}.
Correspondingly, the QCRB can be attained with this interference method
by taking a measurement $\hat{\bm{M}}_{{\rm opt}}$ that makes the
equality $F(\rho_{\theta}\vert\hat{\bm{M}}_{{\rm opt}})\!=\!F_{Q}(\rho_{\theta})$
true. $\hat{\bm{M}}_{{\rm opt}}$ thus represents the optimal measurement
that we would like to find. In what follows, we clarify that the three
conventional measurements as mentioned previously are conditionally
optimal for phase estimation in the MZI. We show the optimal conditions
by identifying that the quantitative statements of the CFI with respect
to the these measurements are equivalent to the expression in Eq.~\eqref{eq:QFI_Jz}. 

\emph{Conventional measurements.---}A TOP particle count measurement
is represented as $\hat{\bm{M}}_{{\rm TOP}}\!=\!\{\vert n_{c},\!n_{d}\rangle\langle n_{c},\!n_{d}\vert\}$,
where the pairs of outcomes $(n_{c},\!n_{d})$ are the number of particles
detected at $c$ and $d$ output ports. The conditional probability
with respect to $(n_{c},\!n_{d})$ is defined by $p(n_{c},\!n_{d}\vert\theta)\!=\!\vert\langle n_{c},\!n_{d}\vert B^{\dagger}\vert\psi_{\theta}\rangle\vert^{2}$.
By identifying $2j\!=\!n_{c}\!+\!n_{d}$ and $2\mu\!=\!n_{c}\!-\!n_{d}$,
we have $p(n_{c},\!n_{d}\vert\theta)=p(j,\!\mu\vert\theta)$, and
further obtain

\begin{equation}
p(j,\!\mu\vert\theta)=\begin{cases}
\bigg|\sum\limits _{\nu=j-\left\lfloor j\right\rfloor }^{j}\!2\thinspace C_{j,\nu}\cos\!\left(\nu\thinspace\theta\right)d_{\nu,\mu}^{j}\!(\frac{\pi}{2})\bigg|^{2},\\
\bigg|\sum\limits _{\nu=j-\left\lfloor j\right\rfloor }^{j}\!2\thinspace C_{j,\nu}\sin\!\left(\nu\thinspace\theta\right)d_{\nu,\mu}^{j}\!(\frac{\pi}{2})\bigg|^{2},
\end{cases}\label{eq:eq:p_twoport}
\end{equation}
where the subscript $\mu\!=\!j,j\!-\!2,j\!-\!4,\cdots\thinspace(j\!-\!1,j\!-\!3,j\!-\!5,\cdots)$
in the upper (lower) expression and $d_{\nu,\mu}^{j}\left(\beta\right)\!=\!\langle j,\!\nu\vert\exp(-i\beta J_{y})\vert j,\!\mu\rangle$
refers to the Wigner rotation matrix. A direct calculation of the
CFI in terms of the TOP measurement with Eq.~\eqref{eq:eq:p_twoport}
suggests that the equality $F(\psi_{\theta}\vert\hat{\bm{M}}_{{\rm TOP}})\!=\!F_{Q}(\psi_{\theta})$
holds when either of the following two conditions is satisfied (see
Appendix A): (a) the amplitude coefficients $C_{j,m}$ of $\vert\psi\rangle$
are real. (b) the true values of $\theta$ are asymptotic to $0$
and $\pi$. This indicates that the TOP measurement is globally optimal
in the whole range of parameter value for real-amplitude symmetric
pure states and locally optimal at points $\theta\!=\!0$ or $\pi$
for complex-amplitude symmetric pure states (see Table \ref{tab:measurements}).
Condition (a) was alternatively obtained in Ref.~\citep{Hofmann2009PRA},
but where there was no statement on condition (b). 

Next we consider the SOP particle count measurement, which is denoted
by $\hat{\bm{M}}_{{\rm SOP}}\!=\!\{\vert n_{c}\rangle\langle n_{c}\vert\}$
in accompany with $n_{c}$ the number of particles counted at $c$
output port. The conditional probability of $n_{c}$ is then given
by $p(n_{c}\vert\theta)\!=\!\vert\langle n_{c}\vert B^{\dagger}\vert\psi_{\theta}\rangle\vert^{2}$.
With the help of the relation $\vert n_{c}\rangle\langle n_{c}\vert\!=\!\sum_{n_{d}=0}^{\infty}\vert n_{c},\!n_{d}\rangle\langle n_{c},\!n_{d}\vert$,
we obtain $p(n_{c}\vert\theta)\!=\!\sum_{n_{d}=0}^{\infty}p(n_{c},\!n_{d}\vert\theta).$
By invoking the Cauchy-Schwarz inequality, one can see that $F(\psi_{\theta}\vert\hat{\bm{M}}_{{\rm TOP}})\!\geq\!F(\psi_{\theta}\vert\hat{\bm{M}}_{{\rm SOP}})$
\citep{Pezze2013PRL}, where the equality holds if and only if 
\begin{equation}
\sqrt{p\left(n_{c},\!n_{d}\vert\theta\right)}=\frac{\lambda}{\sqrt{p\left(n_{c},\!n_{d}\vert\theta\right)}}\frac{d\,p\left(n_{c},\!n_{d}\vert\theta\right)}{d\theta},\label{eq:cauchy-schwarz}
\end{equation}
is satisfied with a nonzero number $\lambda$. Combining Eqs.~\eqref{eq:eq:p_twoport}
and \eqref{eq:cauchy-schwarz}, we find that the equality $F(\psi_{\theta}\vert\hat{\bm{M}}_{{\rm SOP}})\!=\!F(\psi_{\theta}\vert\hat{\bm{M}}_{{\rm TOP}})$
holds only in the asymptotic limits $\theta\!\rightarrow\!0$ and
$\pi$. Therefore this indicates that the SOP measurement can attain
the QCRB for all symmetric pure states at points $\theta\!=\!0$ and
$\pi$ (see Table \ref{tab:measurements}), which is confirmed by
a specific example considered in Ref.~\citep{Pezze2013PRL}.

\begin{table}[b]
\caption{The conditions for achieving the QCRB by three conventional measurements
in the MZI. For a specific type of probe state, condition refers to
phase interval in which the QCRB is saturated. Here $\mathbb{R}_{{\rm S}}$
and $\mathbb{C}_{{\rm S}}$ denote the sets of real- and complex-amplitude
symmetric pure states. Remarkably, for the SOP and TOP measurements,
conditions are also satisfied by replacing $\vert\psi\rangle$ with
the corresponding phase averaged mixed states of Eq.~\eqref{eq:PAS}.
In the last row, the remark of $\langle\Delta\hat{N}^{2}\rangle_{\psi}\!=\!0$
means that $\vert\psi\rangle$ has a definite number of particles.
\label{tab:measurements}}
\begin{ruledtabular}   
\begin{tabular}{cccc}
\thead{Measurement} & $\vert\psi\rangle\in\mathbb{R}_{\rm{S}}$ & $\vert\psi\rangle\in\mathbb{C}_{\rm{S}}$ & Remark\\
\hline
\thead{\tabincell{c}{SOP}}& $\theta\rightarrow 0,\pi$ & $\theta\rightarrow 0,\pi$ & \\
\thead{\tabincell{c}{TOP}}& $0\leq\theta\leq\pi$ & $\theta\rightarrow 0,\pi$ & \\
\thead{\tabincell{c}{PD}}& $0\leq\theta\leq\pi$ & $\theta\rightarrow 0,\pi$ & $\langle\Delta\hat{N}^{2}\rangle_{\psi}\!=\!0$
\end{tabular}   
\end{ruledtabular}
\end{table}

Finally, we discuss the PD measurement described by $\hat{\bm{M}}_{{\rm PD}}\!=\!\{\vert\mu\rangle\langle\mu\vert\}$,
which is also known as the projection measurement with respect to
the observable $\hat{J}_{z}$. Notice that the PD measurement is highly
related to the TOP measurement, since the latter explicitly reduces
to the former when the total particle number of the system is definite.
Therefore one can take the PD measurement to saturate the QCRB when
the probe state belongs to a subclass of symmetric pure states featuring
no fluctuation of particle number, such as, NOON state, twin Fock
state. However, the PD measurement fails for the case in which the
fluctuating particle number presents, since it is insensitive to distinguish
subspaces with different particles numbers. Fortunately, states mostly
used in Ramsey interferometer are absence of the fluctuating number
of particles. Thus, in such circumstances, the PD measurement is applicable.

Otherwise, in optical interferoemtric setting, quantum states with
fluctuating particle number may have coherences between states of
different numbers of particles which are generally not measurable
\citep{Bartlett2007RMP,Hyllus2010PRL,Jarzyna2012PRA}. Now, we consider
a general phase averaged state of the form
\begin{equation}
\rho=\int\frac{d\vartheta}{2\pi}\exp\!\big(\!-i\vartheta\hat{N}\big)\left|\psi\right\rangle \!\left\langle \psi\right|\exp\!\big(i\vartheta\hat{N}\big),\label{eq:PAS}
\end{equation}
where $\left|\psi\right\rangle $ of consideration is the symmetric
pure state. States in expression of Eq.~\eqref{eq:PAS} are vanishing
of the coherence of different particle number states as the consequence
of the lack of a suitable phase reference frame \citep{Bartlett2007RMP,Hyllus2010PRL,Jarzyna2012PRA}.
After the MZI process, $\rho$ evolves to $\rho_{\theta}$. A direct
calculation of the QFI for $\rho_{\theta}$ yields $F_{Q}(\rho_{\theta})\!=\!F_{Q}(\psi_{\theta})$.
Furthermore, by calculating the CFI with respect to the SOP and TOP
measurements, we find that the previous important results for pure
states can be generalized to mixed states in the form of Eq.~\eqref{eq:PAS}.

\begin{figure}[t]
\includegraphics[scale=0.58]{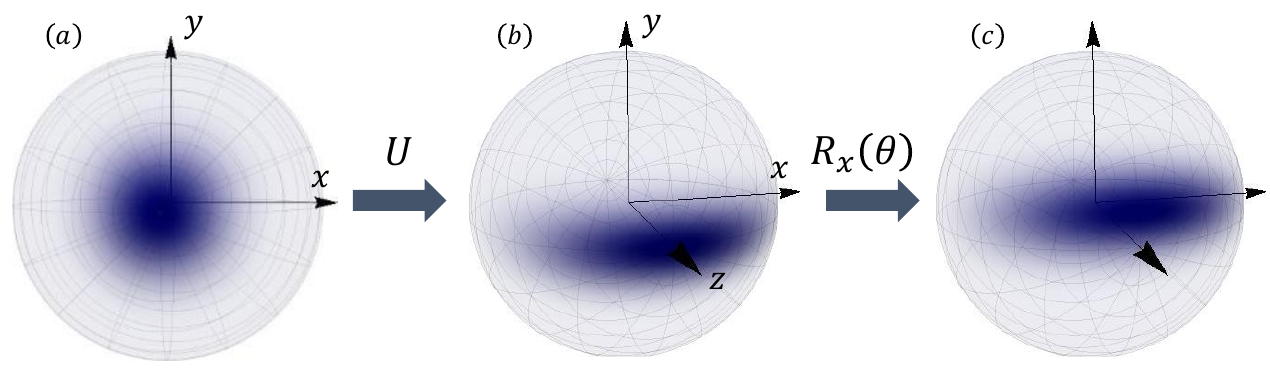}\caption{(Color online) Geometric representation of atomic interferometric
scheme with the TACT state. Illustrated are Husimi distributions for
$N=60$ atoms. \label{fig:Ramsey}}
\end{figure}

\emph{Implementation to Ramsey spectroscopy.---}So far we have presented
experimentally friendly conditions for saturating the QCRB in a linear
MZI. As an example, we apply our results on Ramsey spectroscopy with
detection noise. It is well known that a Ramsey spectroscopy (Fig.~\ref{fig:Ramsey})
is formally analogue to the MZI (Fig.~\ref{fig:MZI}) by replacing
two beam splitters with two Ramsey sequences and phase shift with
free evolution. Below we consider a general case of the PD measurement
with finite resolution $\sigma$, which is modeled by replacing the
ideal probability $p(\mu\vert\theta)$ with 
\begin{equation}
\tilde{p}(\mu\vert\theta)=\!\!\!\sum_{\mu^{\prime}=-N/2}^{N/2}\!\!\!g(\mu-\mu^{\prime})\,p(\mu^{\prime}\vert\theta),\label{eq:noisy_detection}
\end{equation}
where we specify the unbiased Gaussian function $g(\mu\!-\!\mu^{\prime})\!\propto\!\exp[-(\mu\!-\!\mu^{\prime})^{2}/2\sigma^{2}]$
as a resolution function.

Consider a system of $N$ spin-$1/2$ atoms which we express by the
collective angular momentum $\hat{\bm{J}}\!=\!\sum_{i=1}^{N}\hat{\bm{S}}_{i}$
in terms of spin operators $\hat{\bm{S}}_{i}$. The Ramsey interferometric
transformation (from Fig.~\ref{fig:Ramsey}(b) to (c)) corresponds
to the operation $R_{x}\!\left(\theta\right)\!=\!e^{i\theta\hat{J}_{x}}$
in terms of $\theta=\omega\tau$ with $\omega$ atomic frequency and
$\tau$ the interrogation time. We use the two-axis counter-twisted
(TACT) state as the input state of Ramsey interferometer. Suppose
that the atomic ensemble is initially prepared in a coherent spin
state $\vert\psi_{{\rm c}}\rangle_{z}$, i.e., the fully polarized
state along the $z$ axis (Fig.~\ref{fig:Ramsey}(a)). The TACT state
is then generated by time evolution under the nonlinear Hamiltonian
of $\hat{H}_{{\rm t}}\!=\!\chi(\hat{J}_{x}^{2}-\hat{J}_{y}^{2})$
which is known as the two-axis counter-twisting Hamiltonian \citep{Kitagawa1993PRA}.
Several proposals have been presented to simulate this nonlinear Hamiltonian
in various physical systems \citep{Liu2011PRL,Shen2013PRA,Zhang2014PRA,Huang2015PRA}.
Moreover, an additional rotation operation $R_{z}\!\left(\pi/4\right)\!=\!e^{-i\hat{J}_{z}\pi/4}$
is employed to re-orient the squeezing angle of the TACT state so
as to acquire the highest degree of sensitivity enhancement. After
the compound operation $U\!=\!R_{z}\,U_{t}$, the state of the system
becomes (Fig.~\ref{fig:Ramsey}(b))
\begin{equation}
\vert\psi_{{\rm t}}\rangle_{z}=e^{-i\hat{J}_{z}\pi/4}\,e^{-i\chi t(\hat{J}_{x}^{2}-\hat{J}_{y}^{2})}\,\vert\psi_{{\rm c}}\rangle_{z}.\label{eq:TACT}
\end{equation}
By fixing $\chi t\!\sim\!\ln(2\pi N)/2N$ \citep{Kajtoch2015PRA},
one may expect to acquire a near-Heisenberg scaling limit $\sqrt{\upsilon}\thinspace\Delta\theta\!\sim\!1.24/N$. 

\begin{figure}[t]
\includegraphics[scale=0.28]{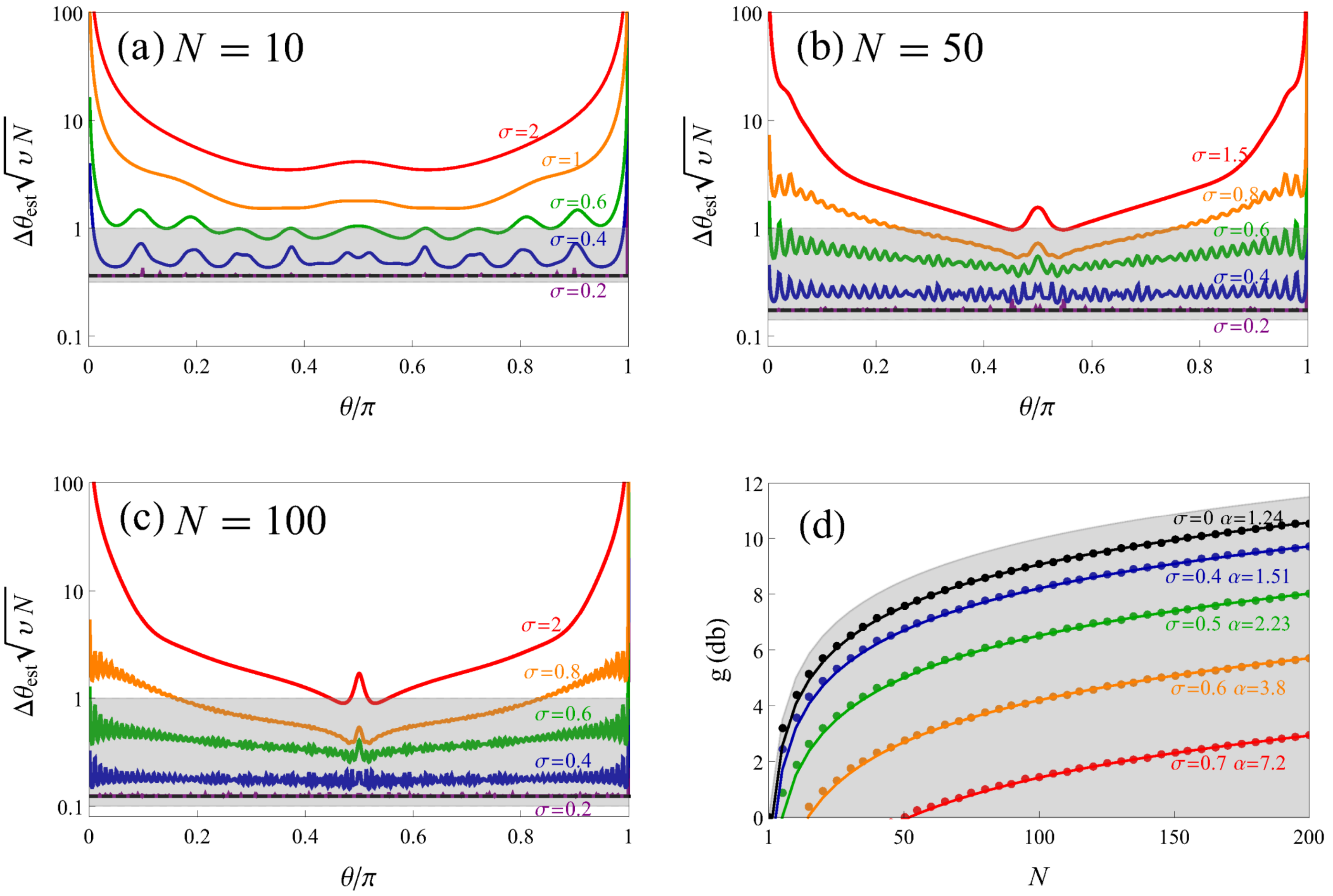}

\caption{(Color online) Phase sensitivity for the TACT states obtained with
a finite-resolution PD measurement (parametrized by $\sigma$). (a-c)
The normalized phase sensitivity $\Delta\theta_{{\rm est}}\sqrt{\upsilon N}$
as a function of $\theta$ for the TACT states of different paticle
number $N$. Different color curves refer to different values of $\sigma$
and the black-dashed horizontal line corresponds to the ideal case
$\sigma=0$. (d) Phase sensitivity gain $g\!\equiv\!-10\log_{10}(\Delta\theta_{{\rm est}}\sqrt{\upsilon N})$
vs $N$ with $\theta\!\sim\!\pi/2N$. Different color filled circles
indicate different values of $\sigma$. Solid lines are the results
of fitting by $g\!\equiv\!-10\log_{10}(\alpha/\sqrt{N})$ where $\alpha\!=\!1$
corresponds to the Heisenberg-limit-scaling sensitivity gain. The
shaded area represents the sub-shot-noise sensitivity region bounded
by shot noise limit and Heisenberg limit. \label{fig:sensitivity}}
\end{figure}

In Fig.~\ref{fig:sensitivity}, we plot the phase sensitivity attained
by the finite-resolution PD measurement. For ideal case $\sigma\!=\!0$,
as shown in Figs.~\ref{fig:sensitivity}(a-c), the near-Heisenberg
scaling sensitivity limit is saturated in the whole phase interval
(see Appendix B). According to previous conclusions, this indicates
a fact that the probe state here, i.e., $\vert\psi_{{\rm t}}\rangle_{x}\!=\!e^{-i\hat{J}_{y}\pi/2}\,\vert\psi_{{\rm t}}\rangle_{z}$,
must be a real symmetric pure state. Nevertheless, for the cases in
presence of detection noise ($\sigma\!\neq\!0$), the phase sensitivity
critically depends on both $\theta$ and $\sigma$. An oscillation
with period of $\pi/N$ takes place for small $\sigma$ and disappears
for more large values of $\sigma$. Phase interval for sub-shot-noise
sensitivity slowly shrinks from the ends of the phase interval towards
the middle as $\sigma$ increases. Notably when a small amount of
$\sigma$ is present, the phase precision becomes significantly worse
at $\theta\!=\!0$ and $\pi$. A similar effect has been found in
Refs.~\citep{Pezze2013PRL,Lucke2011Sci}. Moreover, it proves that
the phase sensitivity becomes more robust against the detection noise
by increasing $N$, which is more pronounced in the regime slightly
departing from $\pi/2$. In order to examine in detail the roles of
the noise power and the size of system on phase sensitivity, we plot
in Fig.~\ref{fig:sensitivity}(d) the phase sensitivity gain as a
function of $N$ with $\theta\!\sim\!\pi/2N$ corresponding to the
first minimum value of the sensitivity in the cases $\sigma\neq0$,
as depicted in Figs.~\ref{fig:sensitivity}(a-c). It is shown that
the amount of $g$ increasingly decreases with the increase of $\sigma$,
but the rate of degradation does not vary with $N$. This means that
for fixed $\sigma$ one can get a higher precision by means of larger
number of probes. For instance, in the case $\sigma\!=\!0.7$, a near
$3\thinspace{\rm db}$ over the shot noise limit can be acquired for
$N\!=\!200$, while no gains obtained for $N\!=\!50$.

In addition, our scheme proves more advantageous than that by means
of the one-axis twisted (OAT) state as $\vert\psi_{{\rm o}}\rangle_{z}\!=\!e^{-i\phi\hat{J}_{z}}\,e^{-i\chi t\hat{J}_{x}^{2}}\,\vert\psi_{{\rm c}}\rangle_{z}$,
where $\phi$ refers to the reorienting angle which rarely depends
on the evolution time $t$ \citep{Kitagawa1993PRA}. It was shown
that the OAT state would provide a phase sensitivity of $\sqrt{\upsilon}\thinspace\Delta\theta\!\sim\!\sqrt{2}/N$
at a platform time $\chi t\!\sim\!1/\sqrt{N}$ \citep{Pezze2009PRL}.
Comparing with the OAT scheme, our scheme offers some attractive features
in precision atomic spectroscopy, such as fixed reorienting angle,
shorter time for creating ideal probe states, and higher measurement
sensitivity. More importantly, our scheme can saturate the QCRB in
the whole range of the phase shift, while it is only valid at two
discrete points $\theta\!=\!0$ and $\pi$ for the OAT case, which
is identified by the fact that its corresponding probe state as $\vert\psi_{{\rm o}}\rangle_{x}\!=\!e^{-i\hat{J}_{y}\pi/2}\,\vert\psi_{{\rm o}}\rangle_{z}$
be a complex symmetric pure state \citep{Strobel2014Sci,Gietka2015PRA}.

\emph{Conclusion.---}We have demonstrated that, under specific conditions,
three different conventional detection methods usually implemented
in interferometric experiments are able to attain the quantum Cram\'er-Rao
phase sensitivity with a Bayesian statistical method. Interestingly,
these conditions are readily met in most of the current experiments
on high precision phase measurement. Therefore, this work may have
practical impact on gravitational wave detection, atomic clock, and
magenetometry and may even be applied for the detection of multipartite
entanglement \citep{Strobel2014Sci,Hauke2016NAT}. 

We thank Xiao-Ming Lu for helpful discussions. This work was supported
by the NKRDP of China (Grant No. 2016YFA0301803) and the NSFC (Grants
No. 91636218). YXH thank the support of the NSF of Zhejiang province
through Grants No. LQ16A040001 and the NSFC through Grants No.11605157.
XGW also acknowledge the support of the NSFC through Grants No. 11475146. 

\bibliographystyle{apsrev4-1}
\bibliography{CFI}

\onecolumngrid

\newpage{}

\appendix

\section*{Appendix A: Analytic solutions of the CFI in terms of the TOP measurement\label{sec:AA}}

  \makeatletter \renewcommand{\theequation}{A\arabic{equation}} \makeatother \setcounter{equation}{0}

In this section, we demonstrate in detail that the equality of $F(\psi_{\theta}\vert\hat{\bm{M}}_{{\rm TOP}})=F_{Q}(\psi_{\theta})$
holds under the following two situations: (a) the expansion coefficients
of probe state on $\vert j,m\rangle$ are real, that is, $\ensuremath{\vert\psi\rangle\in\mathbb{R}_{{\rm S}}}$.
(b) the true value of the phase is asymptotic to $0$ and $\pi$. 

As for the first case, Equation~(3) in the main text then reduces
to 
\begin{equation}
p(j,k\vert\theta)=\begin{cases}
\left[\sum\limits _{\nu=j-\left\lfloor j\right\rfloor }^{j}2\thinspace C_{j,\nu}\cos\left(\nu\thinspace\theta\right)d_{\varsigma,j-k}^{j}\left(\frac{\pi}{2}\right)\right]^{2}, & k\thinspace{\rm for}\thinspace{\rm even}\\
\left[\sum\limits _{\nu=j-\left\lfloor j\right\rfloor }^{j}2\thinspace C_{j,\nu}\sin\left(\nu\thinspace\theta\right)d_{\varsigma,j-k}^{j}\left(\frac{\pi}{2}\right)\right]^{2}, & k\thinspace{\rm for}\thinspace{\rm odd}
\end{cases}\label{eq:A_p1}
\end{equation}
due to the reality of the amplitude coefficient $C_{j,\nu}$ and of
the Wigner rotation matrix $d_{\nu,\mu}^{j}(\frac{\pi}{2})$. Note
that the above expressions are also valid when the amplitude coefficient
$C_{j,\nu}$ contains a phase factor $e^{i\phi_{j}}$ which only depends
on $j$. In this case, we incorporate the phase factor into the basis
vector so as to ensure the expansion coefficient remains real. Thus
by substituting Eq.~\eqref{eq:A_p1} into Eq.~(2) in the main text,
we obtain 
\begin{eqnarray}
F(\psi_{\theta}\vert\hat{\bm{M}}_{{\rm TOP}}) & = & \sum_{j}\sum_{k}\frac{1}{p(j,k\vert\theta)}\bigg(\frac{d\,p(j,k\vert\theta)}{d\theta}\bigg)^{2}\nonumber \\
 & = & 4\sum_{j}\!\!\sum_{k={\rm even}}\!\left[\!\sum\limits _{\nu=j-\left\lfloor j\right\rfloor }^{j}\!\!2\thinspace C_{j,\nu}\thinspace\nu\thinspace\cos\left(\nu\thinspace\theta\right)d_{\nu,j-k}^{j}\!\left(\frac{\pi}{2}\right)\right]^{2}\!\!+4\sum_{j}\!\!\sum_{k={\rm odd}}\!\left[\!\sum\limits _{\nu=j-\left\lfloor j\right\rfloor }^{j}\!\!2\thinspace C_{j,\nu}\thinspace\nu\thinspace\sin\left(\nu\thinspace\theta\right)d_{\nu,j-k}^{j}\!\left(\frac{\pi}{2}\right)\right]^{2}\nonumber \\
 & = & 8\sum_{j}\sum\limits _{\nu=j-\left\lfloor j\right\rfloor }^{j}C_{j,\nu}^{2}\thinspace\nu^{2}\cos^{2}\left(\nu\thinspace\theta\right)+8\sum_{j}\sum\limits _{\nu=j-\left\lfloor j\right\rfloor }^{j}C_{j,\nu}^{2}\thinspace\nu^{2}\sin^{2}\left(\nu\thinspace\theta\right)\nonumber \\
 & = & 8\sum_{j}\sum\limits _{\nu=j-\left\lfloor j\right\rfloor }^{j}C_{j,\nu}^{2}\thinspace\nu^{2},
\end{eqnarray}
where the penultimate equality follows from the following identities
\begin{equation}
\sum_{k={\rm even}}d_{\mu,j-k}^{j}\!\left(\frac{\pi}{2}\right)d_{\nu,j-k}^{j}\!\left(\frac{\pi}{2}\right)=\sum_{k={\rm odd}}d_{\mu,j-k}^{j}\!\left(\frac{\pi}{2}\right)d_{\nu,j-k}^{j}\!\left(\frac{\pi}{2}\right)=\frac{1}{2}\thinspace\delta_{\mu,\nu}.\label{eq:identities}
\end{equation}
It can be verified based on the normalization relation of $\langle\psi\vert B^{\dagger}B\vert\psi\rangle=1$
associating with $\vert\psi\rangle$ being a symmetric pure state.
This is a key ingredient in obtaining the exact solution of the CFI.

In the second situation, Equation.~(3) in the main text can be simplified
in the limit $\theta\rightarrow0$ to 
\begin{equation}
p(j,k\vert\theta)=\begin{cases}
\bigg|\sum\limits _{\nu=j-\left\lfloor j\right\rfloor }^{j}2\thinspace C_{j,\nu}\,d_{\nu,j-k}^{j}(\frac{\pi}{2})\bigg|^{2}, & k\thinspace{\rm for}\thinspace{\rm even},\\
\bigg|\sum\limits _{\varsigma=j-\left\lfloor j\right\rfloor }^{j}2\thinspace\nu\thinspace\theta\thinspace C_{j,\nu}\,d_{\nu,j-k}^{j}(\frac{\pi}{2})\bigg|^{2}, & k\thinspace{\rm for}\thinspace{\rm odd},
\end{cases}\label{eq:A_p2}
\end{equation}
by omitting higher order terms with respect to $\theta$. Substituting
Eq.~\eqref{eq:A_p2} into Eq.~(3) in the main text gives

\begin{eqnarray}
F(\psi_{\theta}\vert\hat{\bm{M}}_{{\rm TOP}}) & = & \sum_{j}\sum_{k={\rm odd}}\frac{1}{p(j,k\vert\theta)}\bigg(\frac{d\,p(j,k\vert\theta)}{d\theta}\bigg)^{2}\nonumber \\
 & = & \sum_{j}\sum_{k={\rm odd}}\frac{\left[\sum\limits _{\mu,\nu=j-\left\lfloor j\right\rfloor }^{j}8\thinspace\mu\thinspace\nu\thinspace\theta\thinspace C_{j,\mu}\thinspace C_{j,\nu}^{\ast}\thinspace d_{j-k,\mu}^{j}\left(\frac{\pi}{2}\right)d_{j-k,\nu}^{j}\left(\frac{\pi}{2}\right)\right]^{2}}{\sum\limits _{\mu,\nu=j-\left\lfloor j\right\rfloor }^{j}4\thinspace\mu\thinspace\nu\thinspace\theta^{2}\thinspace C_{j,\mu}\thinspace C_{j,\nu}^{\ast}\thinspace d_{j-k,\mu}^{j}\left(\frac{\pi}{2}\right)d_{j-k,\nu}^{j}\left(\frac{\pi}{2}\right)}\nonumber \\
 & = & 16\sum_{j}\sum_{k={\rm odd}}\sum\limits _{\mu,\nu=j-\left\lfloor j\right\rfloor }^{j}\mu\thinspace\nu\thinspace C_{j,\mu}\thinspace C_{j,\nu}^{\ast}\thinspace d_{j-k,\mu}^{j}\left(\frac{\pi}{2}\right)d_{j-k,\nu}^{j}\left(\frac{\pi}{2}\right)\nonumber \\
 & = & 8\sum_{j}\sum\limits _{\nu=j-\left\lfloor j\right\rfloor }^{j}\left|C_{j,\nu}\right|^{2}\thinspace\nu^{2},
\end{eqnarray}
where we have again used Eq.~\eqref{eq:identities} in the penultimate
equality. Following the same procedure, the equality of $F(\psi_{\theta}\vert\hat{\bm{M}}_{{\rm TOP}})=F_{Q}(\psi_{\theta})$
can be also obtained in the asymptotic limit $\theta\rightarrow\pi$.

\section*{Appendix B: Bayesian simulation\label{sec:AB}}

  \makeatletter \renewcommand{\theequation}{B\arabic{equation}}\makeatother \setcounter{equation}{0} 

In this section, we consider the Bayesian phase estimation protocol.
Suppose that $\theta_{0}$ is the true value of phase shift to be
estimated for a given parametric density matrix $\rho_{\theta}$ and
$p(\chi\vert\theta)$ represents the conditional probability of the
outcome result $\chi$ of $\hat{\bm{M}}$ depending on $\theta$.
According to the Bayes theorem, the phase probability distribution
is obtained by $p(\theta\vert\chi)=p(\chi\vert\theta)\,p(\theta)/p(\chi)$,
where $p(\theta)$ is the phase probability distribution prior to
the measurement and $p(\chi)$ gives the normalization. After a sequence
of $\upsilon$ independent measurements $\hat{\bm{M}}$, the posterior
distribution $p(\theta\vert\left\{ \chi_{i}\right\} _{i=1}^{\upsilon})$
of the phase shift $\theta$ conditioned on the measurement outcomes
$\left\{ \chi_{i}\right\} _{i=1}^{\upsilon}=\chi_{1},\ldots,\chi_{\upsilon}$
is given by
\begin{equation}
p(\theta\vert\left\{ \chi_{i}\right\} _{i=1}^{\upsilon})=\frac{p(\theta)\,\prod_{i=1}^{\upsilon}p(\chi_{i}\vert\theta)}{\prod_{i=1}^{\upsilon}p(\chi_{i})},
\end{equation}
where the prior knowledge is assumed to be a uniform distribution
as $p(\theta)=1$ and the denominator $\prod_{i=1}^{\upsilon}p(\chi_{i})$
serves as the normalization. The estimator $\theta_{{\rm est}}\left(\left\{ \chi_{i}\right\} _{i=1}^{\upsilon}\right)$
is chosen as the maximum of the posterior distribution $p(\theta\vert\left\{ \chi_{i}\right\} _{i=1}^{\upsilon})$,
which, in the asymptotic limit $\upsilon\rightarrow\infty$, becomes
normally distributed centered around the true value $\theta_{0}$
and with variance $\sigma^{2}=1/\upsilon F(\rho_{\theta}\vert\hat{\bm{M}})$
\citep{Uys2007PRA,Krischek2011PRL,Strobel2014Sci}. Thus this estimation
scheme can saturate the classical Cram\'er-Rao lower bound $\Delta\theta_{{\rm est}}^{2}=[\upsilon F(\rho_{\theta}\vert\hat{\bm{M}})]^{-1}$
in the asymptotic limit of measurements. 

\begin{figure}[tbph]
\includegraphics[scale=0.5]{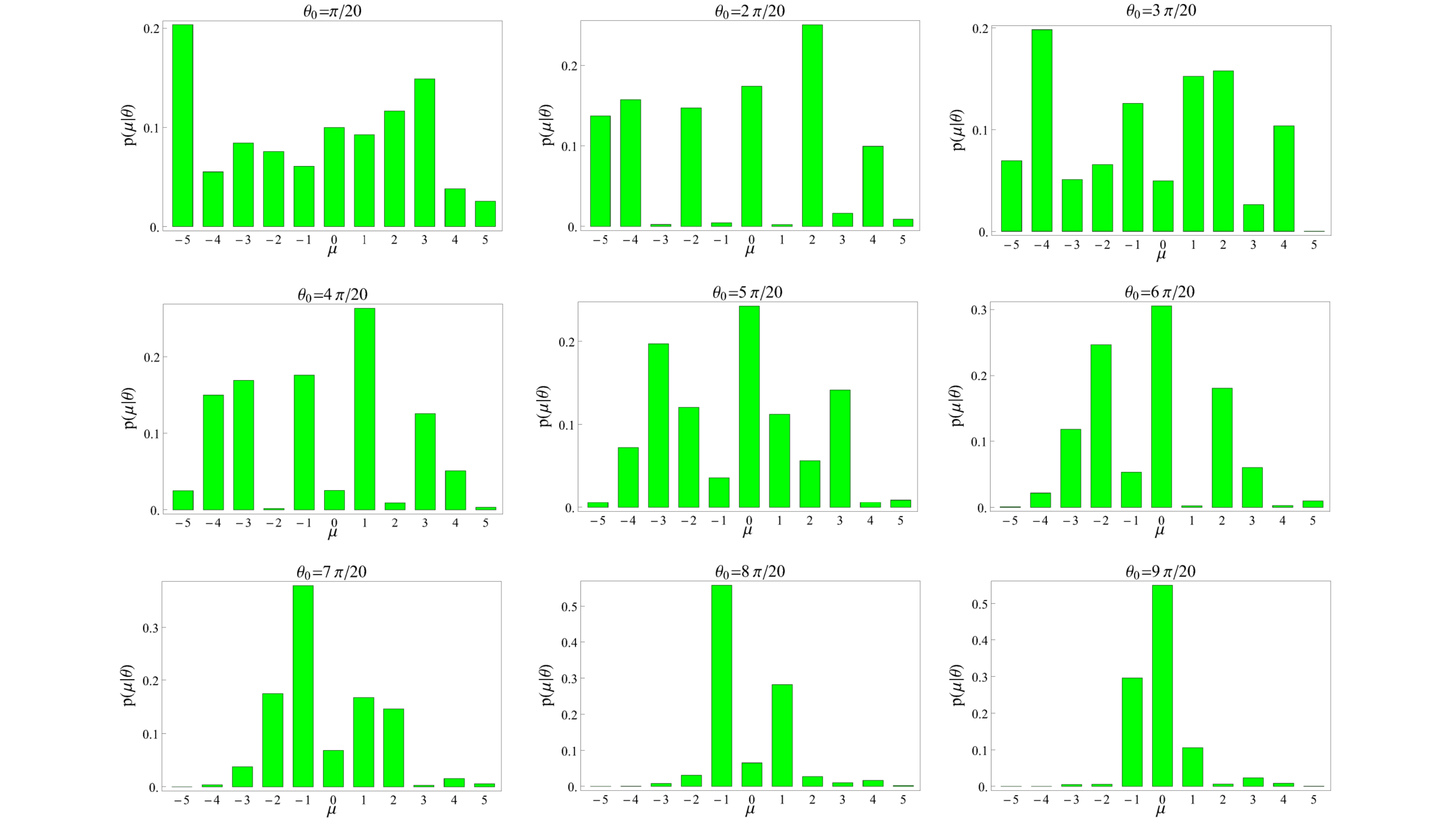}

\caption{(Color online) Probability distributions $p(\mu\vert\theta)$ of outcome
$\mu$ conditioned on different values of $\theta_{0}$ for the TACT
state given by Eq.~(7) in the main text. Here $N=10$ and outcomes
of the PD measurement $\hat{\bm{M}}_{{\rm PD}}$ are thus $\mu=\left\{ 0,\,\pm1,\,\pm2\,\pm3\,\pm4\,\pm5\right\} $.\label{fig:probability}}
\end{figure}

\begin{figure}[tbph]
\includegraphics[scale=0.36]{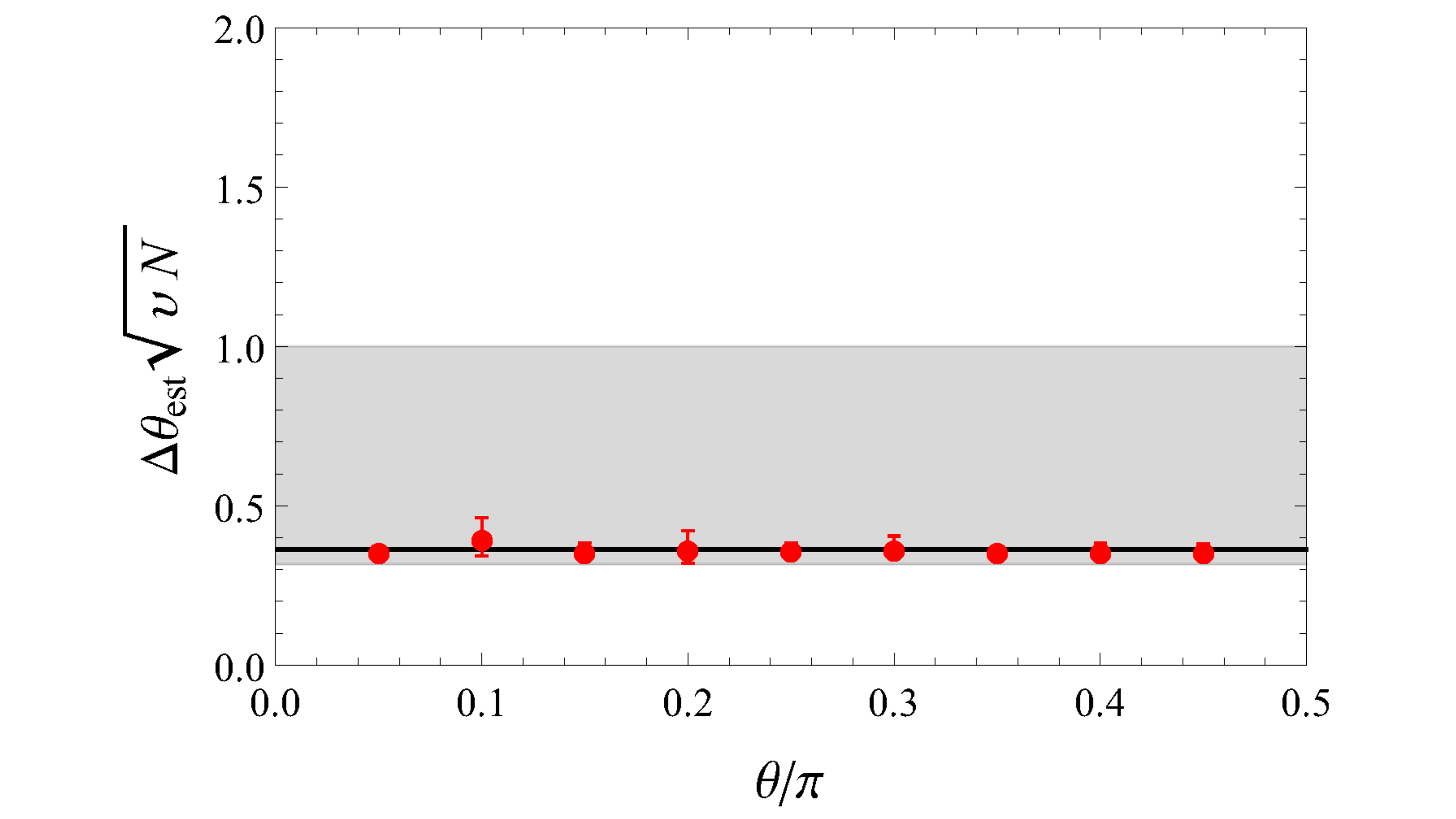}

\caption{(Color online) Normalized phase sensitivity as a function of the phase
shift $\theta$ for the TACT state of particle number $N=10$. The
horizontal black line corresponds to the result of the QCRB. Red-filled
circles are results of numerical simulations of Bayesian analysis
with the number of measurements $\upsilon=100$. The error bars indicate
the standard deviation of a single sequence. The black-solid line
corresponds to the QCRB limited sensitivity. The shaded area represents
the sub-shot-noise sensitivity region bounded by shot noise limit
and Heisenberg limit. \label{fig:Bayesian}}
\end{figure}

Below, we numerically simulate a phase estimation experiment by employing
the Bayesian estimation approach and demonstrate that the PD measurement
$\hat{\bm{M}}_{{\rm PD}}$ is the optimal in the whole phase interval
for achieving the ultimate sensitivity given by the TACT state. Considering
the state $\vert\psi_{{\rm t}}\rangle_{z}$ given in Eq.~(7) in the
main text and the Ramsey interferometry process $R_{x}\left(\theta\right)$
depicted in Fig.~(2) in the main text, the conditional probability
with respect to outcomes $\mu$ of the measurement $\hat{\bm{M}}_{{\rm PD}}$
is given by 
\begin{equation}
p\left(\mu\vert\theta\right)=\left|\langle\mu\vert R_{x}\left(\theta\right)\vert\psi_{{\rm t}}\rangle_{z}\right|^{2}.
\end{equation}
Here, we set the phase shift to $9$ known values $\theta_{0}/\pi=1/20,\,1/10,\ldots,9/20$.
The conditional probabilities for different values of $\theta_{0}$
are plotted in Fig.~\eqref{fig:probability}. To simulate the experiment,
for each $\theta_{0}$, we randomly draw $5000$ repetitions $\mu_{i}$
of this settings and divided them into sequences of length $\upsilon=100$
for each sequence. Using these outcome results, we implement a Bayesian
phase estimation protocol as discussed above. It is known that for
sufficiently large $\upsilon$, the phase posterior probability $p(\theta\vert\left\{ \mu_{i}\right\} _{i=1}^{\upsilon})$
becomes a Gaussian distribution. The phase uncertainty is determined
by the $68\%$ confidence interval around $\theta_{{\rm est}}$ which
corresponds to the maximum of $p(\theta\vert\left\{ \mu_{i}\right\} _{i=1}^{\upsilon})$.
As plotted in Fig.~\eqref{fig:Bayesian}, we show that the phase
sensitivities provided by Bayesian analysis agree with the result
obtained from the QCRB.
\end{document}